\documentclass{PoS}

\def\be{\begin{eqnarray}}
\def\ee{\end{eqnarray}}
\def\el{\nonumber\\}
\ifpdf
  \def\nca#1{\pdfsetcolor{0 0.61 0.87 0}#1\pdfsetcolor{\maincolor}}
  \def\ncb#1{\pdfsetcolor{0 1 1 0}#1\pdfsetcolor{\maincolor}}
\else
  \def\nca#1{\em{#1}}
  \def\ncb#1{\em{#1}}
\fi

\title{Initial guesses for multi-shift solvers }

\ShortTitle{Initial guesses for multi-shift solvers }

\author{\speaker{James Osborn}%
        \\Argonne National Lab \\
        E-mail: \email{osborn@alcf.anl.gov}}

\abstract{
I will present a method for providing initial guesses to a linear solver for systems with
multiple shifts.
This can also be extended to the case of multiple sources each with a
different shift.
}

\FullConference{The XXVI International Symposium on Lattice Field Theory \\
  July 14 - 19, 2008\\
  Williamsburg, Virginia, USA}

\begin{document}

\section{Introduction}

A fundamental part of lattice QCD calculations is the solution of a discretized Dirac
 equation
\be
(D+m) ~ x ~=~ b
\ee
for some source field $b$.  Here $D$ is the Dirac matrix
 (for some choice of discretization),
 $m$ is the quark mass (times the identity) and $x$ is the desired solution.
This is typically solved with methods such as Conjugate Gradient (CG), that
find a solution among the Krylov space $\{b, (D+m)b, (D+m)^2b, ...\}$.

Often it is necessary to solve this equation for several masses against the same source.
This can be done efficiently with a class of Krylov methods that solve
for all shifts at the same time \cite{mskry}.
These methods are made possible because the Krylov spaces for different shifts still
span the same linear space, and the solutions can all be obtained in a single pass
of the algorithm.
The number of iterations required for the solution of all equations is then just the number
needed for the worst conditioned equation (lightest mass).

Since solving the Dirac equation can make up a large part of lattice QCD calculations,
it becomes very important to find ways to reduce the time needed to get a solution.
Often one can use some prior knowledge about similar systems to the one being solved
to obtain initial guesses which, in the case of a single shift, can easily be used to
reduce the number of iterations needed.  The prior knowledge can be from previous solutions
at a lower precision, projection of low eigenmodes (or approximate ones), or solutions of
similar systems with small changes in either the source or the matrix (such as in the
chronological inverter \cite{chrono}).

Unfortunately for systems with multiple shifts the use of the prior information is not as simple.
This is due to the residuals obtained from the guesses not being the same in general.
Here we will present a method that can use this information to produce initial guesses with
a common right hand side so that standard multi-shift Krylov methods can still be used.
The problem of initial guesses is related to the more general problem of solving systems
with multiple shifts each with a different source, which we will also provide an algorithm for.
While the method of initial guesses does provide an improvement in some cases,
 a straightforward implementation
 may in other cases produce initial residuals that are too large to be useful.
We will show examples that demonstrate this breakdown and discuss some possible methods
 to alleviate it.

\section{Multiple shift solvers and initial guesses}

Here we are interested in solving the 
system of $N$ linear equations 
\be
(A+\sigma_i) ~ x_i ~=~ b ~~~ (1 \le i \le N)
\ee
where $A$ is matrix and $\sigma_i$ are shifts of a constant times the identity. 
As mentioned in the introduction, these equations can be solved simultaneously by
using multi-shift Krylov methods that exploit their common Krylov space.
These multi-shift methods form the solutions from the common Krylov space
 $\{b,Ab,A^2b,...\}$.
If one wanted to make use of some initial guesses, $y_i$, for the solutions to reduce
 the number of iterations, the typical thing to do is construct 
\be
r_i ~=~ b - (A+\sigma_i) ~ y_i
\ee
and then solve 
\be
(A+\sigma_i) ~ z_i ~=~ r_i ~.
\ee
The solutions would be then given by $x_i = y_i + z_i$.
However, in general, the new right hand sides, $r_i$, are not collinear, and hence
 they won't share the same Krylov space, preventing the use of multi-shift Krylov
 methods which would solve them all at the same time.

There is a simple form for the guesses that will make the $r_i$ the same.
Take
\be
\label{yig}
y_i ~=~ \left[ \prod_{j\ne i} (A+\sigma_j) \right] w
\ee
for any vector $w$, then one can easily see that the new right hand sides are all equal to
\be
\label{rig}
r_i ~=~ b - \left[ \prod_{j} (A+\sigma_j) \right] w  ~.
\ee
The system can then be solved with standard multi-shift Krylov methods.

The problem now is just to find the best choice for $w$.
Consider first the case of $N=2$ shifts.
Given approximate solutions
\be
v_1 ~\approx~ (A+\sigma_1)^{-1} ~ b \el
v_2 ~\approx~ (A+\sigma_2)^{-1} ~ b
\ee
with corresponding residuals
\be
R_1 ~=~ b - (A+\sigma_1) ~ v_1 \el
R_2 ~=~ b - (A+\sigma_2) ~ v_2
\ee
then a good choice for $w$ could be
\be
w ~=~ (v_1 - v_2)/(\sigma_2 - \sigma_1) ~\approx~ [(A+\sigma_1)(A+\sigma_2)]^{-1} ~ b
\ee
giving
\be
\label{rhs2}
r_1 ~=~ r_2 ~=~ [ (A+\sigma_2) R_1 - (A+\sigma_1) R_2 ] / (\sigma_2 - \sigma_1) ~.
\ee
Note that if $v_1$ and $v_2$ were exact solutions then starting residual 
 would be zero.

For general $N$ the corresponding choice for $w$ would be
\be
\label{wig}
w ~=~ \sum_i \, c_i ~ v_i
\ee
with
\be
c_i ~=~ \prod_{j\ne i} \frac{1}{\sigma_j - \sigma_i} ~.
\ee
Note that the coefficients $c_i$ can become large as one goes to more shifts with
smaller differences.
As we will see later, this can lead to a breakdown of the algorithm if care is not taken
to keep the common residual (\ref{rig}) from growing too large.

\section{Multiple shifts with multiple sources}

It turns out that the problem of initial guesses is a special case of the more
general case of multiple shifts each with a different source, which can also be solved.
The two-source two-shift method was worked out in \cite{mc06}.
Consider the system
\be
(A+\sigma_1) ~ x_1 &=& b_1 \el
(A+\sigma_2) ~ x_2 &=& b_2 ~.
\ee
Now choose guesses $y_k$ such that the residuals are equal
\be
b_1 - (A+\sigma_1) ~ y_1 ~=~ b_2 - (A+\sigma_2) ~ y_2  ~.
\ee 
By equating powers of $A$ we find
\be
y_1 = y_2 = (b_2 - b_1)/(\sigma_2 - \sigma_1) 
\ee
which gives a common starting right hand side of
\be
b_i - (A+\sigma_i) ~ y_i ~=~ [ (A+\sigma_2) ~ b_1 - (A+\sigma_1) ~ b_2 ] / (\sigma_2 - \sigma_1)
\ee
which is just (\ref{rhs2}) with the $b_i$ replaced by $R_i$.

To extend this to arbitrary $N$ we need to find a set of $y_i$ that give a common
residual $r$
\be
b_i - (A+\sigma_i) ~ y_i ~=~ r  ~~~~ (1 \le i \le N) ~.
\ee
This can be solved by setting
\be
y_i ~=~ \sum_{j=0}^{N-2} ~ A^j ~ s_{i,j} ~\equiv~ p_i(A)
\ee
then equating powers of $A$ and solving for the vectors $s_{i,j}$.
One can also solve this by considering the polynomials $q_i(A) = (A+\sigma_i) p_i(A)$
 at the special cases of $A=-\sigma_k$ where the residual $r = b_k$.
This gives the $N$ equations (for fixed $i$)
\be
q_i(-\sigma_k) ~=~ b_i - b_k ~.
\ee
Since $q_i(A)$ is a polynomial of order $N-1$ in $A$ the system is uniquely determined.
The polynomial satisfying these equations is
\be
q_i(A) ~=~ \sum_{k} \left[ \prod_{j \ne k} \frac{A+\sigma_j}{\sigma_j-\sigma_k}
 \right] (b_i - b_k)
\ee
which gives
\be
\label{yimsms}
y_i ~=~ \sum_{k\ne i} \left[ \prod_{j \ne i,k} \frac{A+\sigma_j}{\sigma_j-\sigma_k}
 \right] \frac{b_i - b_k}{\sigma_i - \sigma_k} ~.
\ee

\begin{figure}
\begin{center}
\begin{tabular}{|c|c|c|c|c|c|c|c|c|c|}
\hline
$m_1$ & d & no guess & $N=1$ & $N=2$ & $N=3$ & $N=4$ & $N=5$ & $N=6$ \\
\hline
0.010 & 2 & 683 & 334 ($-3.0$) & 445 ($4.6$) & 488 ($11.0$) & 509 ($16.3$) & 518 ($20.4$) & \nca{523 ($23.3$)} \\
0.010 & $\sqrt{2}$ & 683 & 334 ($-3.0$) & 489 ($5.4$) & 575 ($13.1$) & 635 ($20.3$) & \ncb{677 ($26.8$)} & \ncb{985 ($32.7$)} \\
\hline
0.005 & 2 & 1365 & 666 ($-3.0$) & 892 ($5.8$) & 978 ($13.5$) & 1018 ($19.9$) & \ncb{1039 ($25.3$)} & \ncb{1214 ($29.4$)} \\
0.005 & $\sqrt{2}$ & 1365 & 666 ($-3.0$) & 977 ($6.6$) & 1151 ($15.6$) & \nca{1272 ($23.9$)} & \ncb{1761 ($31.7$)} & \ncb{3066 ($38.8$)} \\
\hline
0.002 & 2 & 3417 & 1668 ($-3.0$) & 2228 ($7.4$) & 2445 ($16.6$) & \nca{2554 ($24.7$)} & \ncb{3965 ($31.6$)} & \ncb{6723 ($37.3$)} \\
0.002 & $\sqrt{2}$ & 3417 & 1668 ($-3.0$) & 2445 ($8.2$) & 2885 ($18.8$) & \ncb{3246 ($28.7$)} & \ncb{7210 ($38.0$)} & \ncb{10000 ($46.8$)} \\
\hline
0.001 & 2 & 6830 & 3353 ($-3.0$) & 4464 ($8.6$) & 4903 ($19.0$) & \ncb{5328 ($28.3$)} & \ncb{10000 ($36.4$)} & \ncb{10000 ($43.3$)} \\
0.001 & $\sqrt{2}$ & 6830 & 3353 ($-3.0$) & 4886 ($9.4$) & 5764 ($21.1$) & \ncb{8949 ($32.3$)} & \ncb{10000 ($42.8$)} & \ncb{10000 ($52.7$)} \\
\hline
\end{tabular}
\caption{Iterations (initial $\log_{10}|r_i|^2$) for guesses formed from approximate solutions.}
\label{tab:soln}
\end{center}
\end{figure}

\section{Initial tests}

To demonstrate the strengths and weaknesses of this method, we have performed some simple tests.
For all tests we are using an even-odd preconditioned ``asqtad'' staggered Dirac matrix
so that we are solving the Hermitian positive definite system
\be
(m_k^2 - D_{eo}D_{oe}) ~ x_k ~=~ b
\ee
where $D_{eo}$ and $D_{oe}$ are the even-odd and odd-even blocks of the Dirac matrix and the shift
 is now the square of the masses.
The source vector $b$ is taken to be a point source.
For simplicity in all tests we used a random gauge field with an average plaquette value of
 around $0.39$ (normalized to 1).
The masses are set to be geometrically spaced, $m_k = m_1  d^{k-1}$, with
 $m_1 \in \{0.01, 0.005, 0.002, 0.001\}$ and $d \in \{2, \sqrt{2}\}$.
The final stopping criterion for the residual is $|r|^2 < 10^{-6}$ (the source is normalized to 1).
While this is a fairly relaxed criterion, it was chosen to keep the number of iterations from
 growing too large at the lightest mass.
There are many factors that can effect the performance of this algorithm, so these tests only
 serve to show the qualitative behavior as more and lighter masses are used.
All work was done in double precision.

The first tests are with initial guesses made from approximate solutions on a $32^4$ lattice.
The approximate solutions were obtained from running multi-shift CG until $|r|^2 < 10^{-3}$. 
These solutions were then used to generate the initial guesses from (\ref{wig}) and (\ref{yig}).
This example is done purely for testing purposes since the residuals $|r_i|^2$ for $i\ge3$
 had already converged to the final precision.
Also since the guesses came from another multi-shift CG, their residuals could have already been
 collinear, which we could have taken advantage of as discussed later.

In figure \ref{tab:soln} we show the results for the approximate solutions.
The ``no guess'' column gives the number of iterations necessary when starting with zero guess.
The $N$ value is the number of equations (shifts) solved simultaneously.  In those columns are
the number of iterations needed before the accumulated residual from the CG reached the stopping
criterion (with a maximum of 10,000 iterations) along with the value of
 $\log_{10}(|r_i|^2)$ for the initial common residual (\ref{rig}) used for the new right hand side.
After the CG stopped, the true residual was calculated for all shifts.
The numbers in orange and red indicate that the true residuals had actually not converged, with
orange for $10^{-6} < |r|^2 < 10^{-5}$ and red for $10^{-5} < |r|^2$.

For all cases that converged, the number of iterations was less with the guess than without.
Remarkably even for a starting residual of $|r_i|^2 \approx 10^{21}$ the residual could be
 reduced to the final precision in fewer iterations.
However as one moves toward more or smaller masses, the initial residual grows very large until
it is no longer possible to reduce it all the way back to the final goal in double precision.

In the second set of tests the guesses were obtained by projection of approximate eigenmodes
 of the preconditioned Dirac matrix.
Here the lattice size was $16^4$.
The low modes were obtained simply by repeated inversions on random vectors with occasional
 Rayleigh-Ritz diagonalization.
The final vectors were still far from the lowest eigenmodes since the smallest
 approximate eigenvalue (Ritz value)
 was still at least 4 times larger than the lowest true eigenvalue.
This was done to give a more difficult test of the algorithm since with exact eigenvalues
 the deflation can be done exactly and the starting residuals are automatically equal.

In figure \ref{tab:eigs} we show the results for the approximate eigenmodes with the same
 conventions as the previous table.
Again we see the same pattern of improved convergence up to the point that the initial residual
 becomes too large to reduce in double precision.
The only exception is at the heaviest mass where the low mode projection is no longer effective
 anyway.
Clearly the method is providing good guesses for the low modes of the system.
The main difficulty then is keeping the initial residual under control so that the solver
 can converge.
Next we will discuss some possible strategies for this.

\begin{figure}
\begin{center}
\begin{tabular}{|c|c|c|c|c|c|c|c|c|c|}
\hline
$m_1$ & d & no guess & $N=1$ & $N=2$ & $N=3$ & $N=4$ & $N=5$ & $N=6$ \\
\hline
0.010 & 2 & 665 & 626 ($0.1$) & 650 ($6.7$) & 677 ($13.0$) & 693 ($18.3$) & 705 ($22.4$) & \ncb{708 ($25.3$)} \\
0.010 & $\sqrt{2}$ & 665 & 626 ($0.1$) & 657 ($7.1$) & 700 ($14.5$) & 740 ($21.5$) & \ncb{772 ($28.0$)} & \ncb{1078 ($33.9$)} \\
\hline
0.005 & 2 & 1328 & 1136 ($0.3$) & 1185 ($8.1$) & 1241 ($15.6$) & \nca{1280 ($22.1$)} & \ncb{1303 ($27.4$)} & \ncb{1605 ($31.5$)} \\
0.005 & $\sqrt{2}$ & 1328 & 1136 ($0.3$) & 1200 ($8.5$) & 1288 ($17.1$) & \ncb{1369 ($25.3$)} & \ncb{1946 ($32.9$)} & \ncb{3263 ($40.0$)} \\
\hline
0.002 & 2 & 3309 & 2157 ($0.4$) & 2268 ($9.5$) & 2402 ($18.5$) & \ncb{2539 ($26.5$)} & \ncb{4776 ($33.4$)} & \ncb{7549 ($39.1$)} \\
0.002 & $\sqrt{2}$ & 3309 & 2157 ($0.4$) & 2291 ($9.7$) & 2479 ($19.6$) & \ncb{3339 ($29.2$)} & \ncb{7274 ($38.4$)} & \ncb{10000 ($47.0$)} \\
\hline
0.001 & 2 & 6563 & 3627 ($0.5$) & 3800 ($10.1$) & 4028 ($20.1$) & \ncb{5969 ($29.3$)} & \ncb{10000 ($37.4$)} & \ncb{10000 ($44.3$)} \\
0.001 & $\sqrt{2}$ & 6563 & 3627 ($0.5$) & 3805 ($10.3$) & \nca{4099 ($21.3$)} & \ncb{8618 ($32.2$)} & \ncb{10000 ($42.6$)} & \ncb{10000 ($52.5$)} \\
\hline
\end{tabular}
\caption{Iterations (initial $\log_{10}|r_i|^2$) for guesses formed from approximate eigenmodes.}
\label{tab:eigs}
\end{center}
\end{figure}

\section{Variations}

For the problem of choosing initial guesses with the same right hand side, there
are several possible strategies for choosing the vector $w$ used in (\ref{wig}).
One possibility is to globally optimize for $w$ from 
$r = b - (A+\sigma_1)\ldots(A+\sigma_N) w$
among some given search space of vectors.
This can be done by either minimizing the norm of the residual or by projecting out
 the search space from the residual.
An alternative is to individually optimize each equation separately,
 $R_k = b - (A+\sigma_k) v_k$,
then apply the multi-source multi-shift algorithm to get the initial guesses.
In practice the latter seems to give better guesses, though global minimizations control
the residual better.  By interpolating between the two one can then find the best compromise
that still leads to a solution.  Of course at some point this will be no better than an initial
guess of zero.

Another variation is given by the observation that the starting residuals don't have to be equal,
but merely collinear.  Thus we can add arbitrary scale factors to the $b_i$ in (\ref{yimsms}).
This is especially useful if the trial guesses are obtained from another run using, e.g., CG.
Here the residuals would be collinear in exact precision, and in finite precision may be close,
but not exact.  Restarting with the appropriate scale factors could give a large improvement
in this case.

\section{Conclusions}

We have presented a method for solving systems with multiple 
 sources each with a different shift.
The main motivation was to provide initial guesses to 
 multi-shift solvers, though it could be useful in other contexts as well.
When used for initial guesses we found that even though the initial residuals may be large,
 the convergence is still typically faster as long as convergence can
 still be reached.
The method breaks down at some point when going to more and/or smaller shifts.
This can be remedied at the expense of using a worse initial guess, which may still
 reduce the number of iterations overall in some cases.
A better solution to this problem may require projecting out the high eigenmodes of the residual
while preserving the low modes of the guesses.


\begin{thebibliography}{99}

\bibitem{mskry}
  R.~W.~Freund,
  \emph{Solution of shifted linear systems by quasi-minimal residual iterations},
  in \emph{Numerical Linear Algebra},
  L.~Reichel, A.~Ruttan and R.S.~Varga (eds.),
  1993;
  U.~Gl\"assner, S.~G\"usken, T.~Lippert, G.~Ritzenh\"ofer, K.~Schilling and A.~Frommer,
  \emph{How to compute Green's functions for entire mass trajectories within Krylov solvers},
  {\tt hep-lat/9605008};
  A.~Bori\c{c}i,
  \emph{Krylov subspace methods in lattice QCD},
  SCSC report TR-96-27;
  B.~Jegerlehner,
  \emph{Krylov space solvers for shifted linear systems},
  {\tt hep-lat/9612014}.


\bibitem{chrono}
  R.C.~Brower, T.~Ivanenko, A.R.~Levi and K.N.~Orginos,
  \emph{Chronological inversion method for the Dirac matrix in hybrid Monte Carlo},
  \emph{Nucl. Phys. B} {\bf 484} (1997) 353.


\bibitem{mc06}
  M.A.~Clark,
  \emph{The rational hybrid Monte Carlo algorithm},
  Ph.D. Thesis, University of Edinburgh (2006).

\end{thebibliography}
\end{document}